\documentclass[a4paper,11pt]{article}

\usepackage{jcappub} 
\usepackage[T1]{fontenc} 

\usepackage{dsfont} 
\usepackage{graphicx}
\usepackage{dcolumn}
\usepackage{bm}
\usepackage[colorlinks=true]{hyperref}
\usepackage[mathlines]{lineno}
\usepackage{aas_macros}
\usepackage{color}
\usepackage{bbm}
\usepackage{multirow}
\usepackage{siunitx}
\usepackage{subcaption}
\usepackage{booktabs}
\usepackage{nicematrix}
\usepackage{amsmath}
\usepackage{caption}
\usepackage{xparse} 
   
\newcommand{\dcom}{\ensuremath{d_\text{M}}}

\newcommand{\Ok}{\ensuremath{\mathcal{O}_k}}

\NewDocumentCommand{\Omm}{o}{\ensuremath{\Omega_{\mathrm{m}\IfNoValueF{#1}{,#1}}}}
\NewDocumentCommand{\Omde}{o}{\ensuremath{\Omega_{\mathrm{de}\IfNoValueF{#1}{,#1}}}}
\NewDocumentCommand{\Oml}{o}{\ensuremath{\Omega_{\Lambda\IfNoValueF{#1}{,#1}}}}
\NewDocumentCommand{\Omk}{o}{\ensuremath{\Omega_{{k}\IfNoValueF{#1}{,#1}}}}

\newcommand{\vb}[1]{\ensuremath{\boldsymbol{#1}}} 
\newcommand{\vect}[1]{\ensuremath{\vb{#1}}} %
\newcommand{\mat}[1]{\ensuremath{{\bm{\mathsf{#1}}}}} 
\newcommand{\rd}{\ensuremath{r_\text{d}}}

\newcommand{\DD}{\ensuremath{\mathcal{D}}}

\newcommand{\ones}{\ensuremath{\mathds{1}}}

\usepackage{scalerel}
\usepackage{tikz}
\usetikzlibrary{svg.path}

\definecolor{orcidlogocol}{HTML}{A6CE39}
\tikzset{
  orcidlogo/.pic={
    \fill[orcidlogocol] svg{M256,128c0,70.7-57.3,128-128,128C57.3,256,0,198.7,0,128C0,57.3,57.3,0,128,0C198.7,0,256,57.3,256,128z};
    \fill[white] svg{M86.3,186.2H70.9V79.1h15.4v48.4V186.2z}
                 svg{M108.9,79.1h41.6c39.6,0,57,28.3,57,53.6c0,27.5-21.5,53.6-56.8,53.6h-41.8V79.1z M124.3,172.4h24.5c34.9,0,42.9-26.5,42.9-39.7c0-21.5-13.7-39.7-43.7-39.7h-23.7V172.4z}
                 svg{M88.7,56.8c0,5.5-4.5,10.1-10.1,10.1c-5.6,0-10.1-4.6-10.1-10.1c0-5.6,4.5-10.1,10.1-10.1C84.2,46.7,88.7,51.3,88.7,56.8z};
  }
}

\newcommand\orcid[1]{\href{https://orcid.org/#1}{\mbox{\scalerel*{
\begin{tikzpicture}[yscale=-1,transform shape]
\pic{orcidlogo};
\end{tikzpicture}
}{|}}}}

\usepackage{scalerel}
\usepackage{tikz}
\usetikzlibrary{svg.path}
\definecolor{orcidlogocol}{HTML}{A6CE39}
\tikzset{
  orcidlogo/.pic={
    \fill[orcidlogocol] svg{M256,128c0,70.7-57.3,128-128,128C57.3,256,0,198.7,0,128C0,57.3,57.3,0,128,0C198.7,0,256,57.3,256,128z};
    \fill[white] svg{M86.3,186.2H70.9V79.1h15.4v48.4V186.2z}
                 svg{M108.9,79.1h41.6c39.6,0,57,28.3,57,53.6c0,27.5-21.5,53.6-56.8,53.6h-41.8V79.1z M124.3,172.4h24.5c34.9,0,42.9-26.5,42.9-39.7c0-21.5-13.7-39.7-43.7-39.7h-23.7V172.4z}
                 svg{M88.7,56.8c0,5.5-4.5,10.1-10.1,10.1c-5.6,0-10.1-4.6-10.1-10.1c0-5.6,4.5-10.1,10.1-10.1C84.2,46.7,88.7,51.3,88.7,56.8z};
  }
}

\definecolor{deepmagenta}{rgb}{0.8, 0.0, 0.8}
\definecolor{ballblue}{rgb}{0.13, 0.67, 0.8}
\definecolor{darkteal}{rgb}{0.0, 0.45, 0.46}
\definecolor{RedWine}{rgb}{0.743,0,0}

\arxivnumber{2601.20293} 
\title{Model independent test of the FLRW metric and the curvature in light of DESI DR2}

 \author[a]{Cléa Millard\orcid{0009-0004-4642-9782},}
\author[a]{Benjamin L'Huillier\orcid{0000-0003-2934-6243},}
\author[b]{Marian Douspis\orcid{}}

\affiliation[a]{Department of Physics and Astronomy, Sejong University, 05006 Seoul, Republic of Korea}
\affiliation[b]{Institut d'Astrophysique Spatiale, Université Paris Saclay \& CNRS UMR 8617,  F-91405 Orsay Cedex, France}

\emailAdd{ccm@sju.ac.kr}
\emailAdd{benjamin@sejong.ac.kr}
\emailAdd{marian.douspis@universite-paris-saclay.fr } 

\abstract{We perform a data-driven test of the FLRW metric and the flatness of the Universe, independently of any Dark Energy model, and in light of the latest DESI DR2 results.
We use Pantheon+ and DES~Dovekie SNIa data to reconstruct the distance modulus, dimensionless comoving distance and Hubble parameter, using an iterative smoothing algorithm. 
Then, combining the various reconstructions with the recent BAO measurements from DESI DR2, we perform the $\mathcal{O}_k$  diagnostic, a litmus test of the FLRW metric and the flatness of the Universe.
We obtain robust results that do not depend on Dark Energy models and test some of the underlying hypotheses of the concordance model. 
We find that when the reconstructed $\mathcal{O}_k$ diagnostic is consistent with the FLRW metric, then the median value of $\Omk[0]$ over all reconstructions that provide an improved fit relative to the flat $\Lambda$CDM model are:${\Omega}_{k,0}^\text{med} = 0.045 ^{+0.045}_{-0.081}\pm 0.038$ for the Pantheon+ \& DESI DR2 data combination, ${\Omega}_{k,0}^\text{med} = 0.095 ^{+0.063}_{-0.136} \pm 0.063$ for the same data but with the Pantheon+ SNIa cut at redshift $z=1.13$, which is the maximum redshift of the DES~Dovekie data, and ${\Omega}_{k,0}^\text{med} = -0.102^{+0.099}_{-0.005}\pm 0.043$ for DES~Dovekie \& DESI DR2. The first uncertainties correspond to the spread in $\Omk[0]$ over all reconstructions, followed by the median 1$\sigma$ error. Our results are consistent with flatness and Planck 2018 within 3$\sigma$.}
 
\arxivnumber{2601.20293}

\begin{document}
\maketitle
\flushbottom

\newpage

\section{Introduction}
\label{sec:intro}
The current concordance model of cosmology, the $\Lambda$CDM model, describes a flat universe with an energy budget dominated by dark energy (DE) in the form of the cosmological constant $\Lambda$ and cold dark matter (CDM). Since the observation of accelerated expansion using Type Ia supernovae (SNIa) \cite{Riess_1998,Perlmutter_1999}, the constraints on the model have been improved by several other probes, among them the cosmic microwave background (CMB, Planck \cite{Planck18}) or the baryon acoustic oscillations (BAO, \cite{desidr2}). However, despite its success, some tensions have arisen in the $\Lambda$CDM model, especially in determining the value of the Hubble constant $H_0$. To this day, the late-time direct measurements of $H_0$ remain in tension with the early-time inferences (for a given model e.g. $\Lambda$CDM) at a $>3\sigma$ level \cite{Di_Valentino_2025}. To resolve this tension, numerous alternative DE models have been suggested \cite{tension_review_2021} but with limited success. Recently, the second data release of the Dark Energy Spectroscopic Instrument (DESI) hinted significantly at a dynamical DE equation of state, shaking the cosmological constant model from $\Lambda$CDM. Alternatively, an early-Universe interpretation with dark acoustic oscillations (DAO), which may bias BAO peak measurements if present near the BAO scale, was suggested in \cite{garny_2025, garny_2025bis} to address the DESI anomaly.

In addition to parametric fits of alternative models, model-independent reconstructions of the expansion history of the Universe allow to directly test the consistency of the data with the assumptions underlying these models. In particular, one of the most fundamental assumptions underlying the $\Lambda$CDM model is that the Universe is homogeneous and isotropic. This assumption is encoded into the Friedmann--Lemaître--Robertson--Walker (FLRW) metric, in which the expansion history of the Universe is governed by the following equation:
    \begin{equation}
        \label{eq:dimensionless_hubble_parameter}
        h(z) = \sqrt{ \Omm[0](1+z)^3 + \Omk[0](1+z)^2 + \Omde[0] \exp  \left( 3\int_0^z \frac{1+w(z')}{1+z'} \mathrm{d}z'\right)}
    \end{equation}
    where $h$ is the dimensionless Hubble parameter, $w(z) = P/ \rho$ is the equation of state of DE, $\Omm[0], \Omk[0], \Omde[0] $ are the density parameter of matter, curvature, and DE today. Then, the dimensionless comoving distance of an astronomical object is, for all signs of $\Omk[0]$:
    \begin{equation}
        \label{eq:general_dimensionless_comoving_distance}
         \mathcal{D}(z) = \frac{1}{\sqrt{-\Omk[0]}} \sin \left( \sqrt{-\Omk[0]} \int_0^z \frac{dz'}{h(z')} \right),
    \end{equation}
    which is related to the luminosity distance $d_{\mathrm{L}}(z)$ and the comoving distance $\dcom(z)$ by:
    \begin{equation}
        \label{eq:general_comoving_distance}
        \frac{c}{H_0} \mathcal{D}(z) = \dcom(z) = \frac{d_{\mathrm{L}}(z)}{1+z}.
    \end{equation}
    The distance modulus of type Ia supernovae is directly linked to the luminosity distance by:
    \begin{equation}
        \label{eq:distmod}
        \mu(z) = 5 \log_{10} \left( \frac{d_{\mathrm{L}}(z)}{\SI1{Mpc}} \right) + 25 = m_\text{B}-M_\text{B},
    \end{equation}
    where $m_\text{B}$ and $M_\text{B}$ are the apparent and absolute magnitudes of SNIa, so that:
    \begin{equation}
    \label{eq:comoving_dist_to_distmod}
        \mathcal{D}(z) = \frac{H_0} {c}\frac{10^{\frac{\mu (z)}{5}-5}}{1+z} \times \SI{1}{Mpc}.
    \end{equation}
    In the case of a flat universe ($\Omk[0] = 0$), the dimensionless comoving distance becomes:
    \begin{equation}
        \label{eq:flat_dimensionless_comoving_distance}
        \mathcal{D}(z) = \int_0^z \frac{dz'}{h(z')}.
    \end{equation}
    
   The transverse and radial modes of the BAO data provide the ratio of the comoving distance $\dcom$ and of the Hubble distance $d_{\mathrm{H}}(z) = c/H(z)$ to the sound horizon at drag epoch $\rd$.

    In the search for beyond $\Lambda$CDM physics, it is important to establish null tests of these underlying assumptions. The goal of model-independent null test of the FLRW metric is to detect deviations from FLRW that would indicate a break from the usual homogeneity and anisotropy assumption, and this without assuming any DE model.
    In this work, we  use the $\Ok$ diagnostic \cite{Ok_diag,Ok_diag_bis}, a litmus test of the FLRW metric and the spatial curvature. We  reconstruct the $\Ok$ diagnostic using a combination of BAO data from DESI DR2 and model-independent reconstructions of the dimensionless distance of SNIa from Pantheon+ or the Dark Energy Survey (DES) Y5. The methodology is taken from \cite{ben_arman2017,arman_ben2018,ben2025}. Additional model-independent null tests of the FLRW metric have been performed using SNIa and BAO data \cite{Montanari_2017,Du_2025, Dias_2025}, BAO DESI DR2 data alone \cite{Dinda_2025}, and strong-lensing time-delay observations \cite{Collet_2019}. The advantage of our combined dataset, model-independent approach, is that the results of the litmus test do not depend on the values of $H_0$ and $r_\mathrm{d}$. We also introduce a p-value test to define which reconstructions of the $\Ok$ diagnostic are consistent with the FLRW metric, and for which the value of $\Ok$ can be interpreted as $\Omk[0]$.
    The paper is organised as follows: Section \ref{sec:data_method} describes the methods and data used to construct the $\Ok$ test, the p-value test and the results are presented in Section \ref{sec:results} and discussion takes place in Section \ref{sec:conclu}.
 
\section{Data and method}
\label{sec:data_method}

    \subsection{Litmus test}
    \label{sec:method_litmus_test}

    The $\Ok$ diagnostic \cite{Ok_diag,Ok_diag_bis} is a litmus test for the FLRW metric. 
    It is defined as :
    \begin{subequations}
    \begin{align}
        \label{eq:Ok}
        \mathcal{O}_k(z) & = \frac{\Theta^2(z) - 1}{\mathcal{D}^2(z)},\\
        \label{eq:Theta}
        \Theta(z) & = h(z) \mathcal{D}'(z) = \sqrt{1+\Omk[0] \DD^2(z)},
    \end{align}
    \end{subequations} 
    where $'$ denotes differentiation with respect to redshift. 
    In an FLRW universe, $\Ok$ is constant such that $ \Ok(z) \equiv \Omk[0]$, so we expect $\Ok = 0$ if this universe is also flat. 
    If a deviation from $\mathcal{O}_k(z) = \mathrm{constant}$ is observed, it is an indication that the data might not be consistent with the FLRW metric. 
    
    As pointed out by Refs.~\cite{ben_arman2017,arman_ben2018,ben2025}, the quantity $c/(H_0\rd)$ can be computed using two methods:
    \begin{subequations}
    \label{eq:c_H0rd}
    \begin{align}
        \label{eq:c_H0rd_methodA}
        \frac{c}{H_0 \rd} &= \frac{1}{\mathcal{D}(z)} \frac{\dcom(z)}{\rd} \\
        \label{eq:c_H0rd_methodB}
        \frac{c}{H_0 \rd} &= h(z) \frac{c}{H(z) \rd}.
        \intertext{In a flat FLRW universe, equation \eqref{eq:c_H0rd_methodB} becomes:}
        \frac{c}{H_0 \rd} & = \frac{1}{\mathcal{D}'(z)} \frac{d_{\mathrm{H}}(z)}{\rd},
        \label{eq:c_H0rd_methodB_flat}
    \end{align}
    \end{subequations}
    and the parameter $\Theta$ can be re-written as the ratio between the two methods:
    \begin{equation}
        \label{eq:Theta_combined}
        \Theta(z) = \frac{\dcom(z)/\rd}{d_{\mathrm{H}}(z)/\rd} \frac{\mathcal{D}'(z)}{\mathcal{D}(z)}.
    \end{equation}
  Using equations~\eqref{eq:Ok} and \eqref{eq:Theta_combined}, we can reconstruct the $\Ok$ diagnostic using BAO data (providing $d_\mathrm{M}/r_\mathrm{d}$ and $d_\mathrm{H}/r_\mathrm{d}$) and model-independently reconstructed $\mathcal{D}$ and $\mathcal{D}'$ from SNIa data, at the redshift of the BAO.

    \subsection{Data}
    \label{sec:data}
    
    To construct the $\Ok$ diagnostic we use the transverse and radial modes of DESI~DR2 BAO data within the redshift range covered by the SNIa data. These quantities are obtained from the clustering of galaxies measured by DESI, through the analysis of the two-point correlation function of the galaxy distribution. The BAO data points and their correlations are taken from \cite{desidr2}, Table IV. The redshift bins are uncorrelated, however, we take into account the correlation between the transverse and radial modes. We note that, while the measurement $d_\mathrm{M}/r_\mathrm{d}$ does not assume flatness, a flat FLRW universe is assumed in the whole BAO pipeline.
    Additionally, we use two compilations of SNIa: Pantheon+ and DES~Dovekie, from which we will reconstruct $\mathcal{D}$ and $\mathcal{D}'$. The Pantheon+ sample consists of 1550 spectroscopically confirmed SNIa, spanning a redshift 0.001 to 2.26~\cite{Pantheon+}. In contrast, the latest DES~Dovekie compiles 1820 SNIa within $0.025<z<1.144$ \cite{desd_dov}, a re-analysis of the DES~Y5 sample \cite{desy5}. For both datasets, the standardisation of the SNIa assumes the $\Lambda$CDM model. However, as it was established in \cite{Koo_2020}, the constraints on the light-curve hyper-parameters are independent of the assumed cosmological model.
    
    When combining the DESI~DR2 BAO data with Pantheon+, we use the five BAO points comprised in $0.510<z<1.484$, as to not exceed the maximum redshift of the SNIa compilation. For the same reason, we only use three BAO data points such as $0.510<z<0.934$ when combining them with DES~Dovekie.
    
    \subsection{Iterative smoothing}
    \label{sec:method_its}

    Since the BAO data are only available at a few specific redshifts $z_\mathrm{BAO}$, we need to estimate the dimensionless comoving distance $\mathcal{D}$ and its derivative $\mathcal{D}'$ at those $z_\mathrm{BAO}$ to combine the two datasets into eq.~\eqref{eq:Theta_combined}. The resulting $\Ok$ diagnostic \eqref{eq:Ok} is then only estimated at $z_\mathrm{BAO}$.
    The non-parametric method of iterative smoothing \cite{arman2006,arman2007,ben2025} allows us to estimate $\mu(z)$, $\mathcal{D}(z)$ and $\mathcal{D}'(z)$ at any arbitrary redshift, and particularly at the redshift of the BAO. It is also independent of any DE model, allowing a model-independent estimation of the value of $\Ok$ at the BAO redshifts.
    The distance modulus $\hat{\mu} (z)$ is reconstructed directly  from an initial guess $\hat{\mu}_0(z)$, using a data set $ \left\{ z_i,\mu_i \right\}$ with a covariance matrix $\mat{C}_\mathrm{SNIa}$. At each iteration $n+1$, a Gaussian kernel with a smoothing scale $\Delta$ smooths the residuals $\vect{\delta \mu}_{n}$ between the data and the reconstruction $n$:
    \begin{equation}
     \label{correlated_reco}
        \hat{\mu}_{n+1}(z) = \hat{\mu}_n(z) + \frac{ \vect{W}^\top(z)\, \mat{C}_\mathrm{SNIa}^{-1} \,\vect{\delta \mu}_n}
        {\vect{W}^\top(z)\, \mat{C}_\mathrm{SNIa}^{-1} \, \ones},
    \end{equation} 
    where the weights $\vect{W}$ and the residuals $\delta \mu_n$ are defined as:
    \begin{subequations}
    \begin{align}
    \label{weight_and_residual}
        \left[\vect{W}(z)\right]_i &= \exp \left( - \frac{\ln^2 \left( \frac{1+z}{1+z_i}\right)}{2\Delta^2}\right), \\
        \left[\vect{\delta \mu}_{n}\right]_i &= \mu_i - \hat{\mu}_n(z_i), \\
        \ones^\top & = (1,\dots,1).
    \end{align} 
    \end{subequations}
     The smoothing scale determines the weight that a data point will have in the smoothing depending on its distance from the test point $z$. The larger $\Delta$, the more influence the distant data point will have on the smoothing.  
     We  use $\Delta = 0.3$, in accordance with previous studies \cite{Ok_diag_bis,ben_arman2017,arman_ben2018,arman2006,arman2007,ben_arman2018,hanwool2020,hanwool2021,hanwool2022}.  Regardless of the initial guess, successive reconstructions will converge toward the same result~\cite{ben_arman2017,arman_ben2018,arman2006,arman2007}. The final collection of reconstructions forms a non-exhaustive set of plausible expansion histories. From the collection of reconstructed distance moduli $\{\widehat{\mu}_n(z)\}$, where $n$ indexes the reconstructions, we can obtain the collection of reconstructed dimensionless distances $\{\widehat{\mathcal{D}}_n(z)\}$ through eq.~\eqref{eq:comoving_dist_to_distmod}. Their smoothed derivative $\{\widehat{\mathcal{D}}'_n(z)\}$ is also obtained by smoothing, as described in \cite{ben2025}. Since eq~\eqref{eq:comoving_dist_to_distmod} requires a value for $H_0$ and the $\{\widehat{\mu}_n(z)\}$ are reconstructed up to an additive constant, we normalize $\{\widehat{\mathcal{D}}_n(z)\}$ and
     $\{\widehat{\mathcal{D}}'_n(z)\}$ as described in Annex~\ref{sec:normalisation_method}, to obtain the normalised collections of $\{\mathcal{D}_n(z)\}$ and
     $\{\mathcal{D}'_n(z)\}$, which we will use for the $\Ok$ diagnostic. This normalisation neutralises any residual effects from the choice of $M_B$, and spares us from assuming $H_0$. 
      
    \subsection{\texorpdfstring{$\chi^2$}{chi-squared} selection criteria}
    \label{sec:selection_criterion}

    As explained in section \ref{sec:method_litmus_test}, we can now combine $\{\mathcal{D}_n\}$ and $\{\mathcal{D}'_n\}$ with the BAO measurements to calculate the $\Ok$ diagnostic.
    For each reconstruction, three different $\chi^2$ can be considered : 
    \begin{enumerate}
        \item The $\chi^2$ to the SNIa data alone:
        \begin{subequations}
        \begin{align}
        \label{eq:chi2_SNIa}
            &\bm{\delta \mu} = \widehat{\mu_n}(\bm{z}_{\mathrm{SNIa}}) - \bm{\mu},\\
            &\chi^2_{\mathrm{SNIa}} = \bm{\delta \mu}^\top
             \mat{C}_{\mathrm{SNIa}}^{-1}
             \bm{\delta \mu},
         \end{align}
        \end{subequations}
        where $\bm{z}_{\mathrm{SNIa}}$ and $\bm{\mu}$ are of dimension $N_\mathrm{SNIa}$, the total number of SNIa in the sample.
        
        \item The $\chi^2$ to the SNIa and the transverse mode of the BAO $\dcom/\rd$:
        \begin{subequations}
        \begin{align}
        \label{eq:chi2_SNIa_dm}
            &\bm{\delta y_\mathrm{M}} = \theta_{\mathrm{M},n} \mathcal{D}_n(\bm{z}_{\mathrm{BAO}}) - \bm{\frac{\dcom}{\rd}},\\
            &\chi^2_{\mathrm{SNIa} + \dcom} = \chi^2_{\mathrm{SNIa}} 
                                            + \bm{\delta y_\mathrm{M}}^\top
                                              \mat{C}_{d_{\mathrm{M}/\rd}}^{-1}
                                             \bm{\delta y_\mathrm{M}},
        \end{align}
        \end{subequations}
        where 
        \begin{equation}
        \label{eq:theta_Mn}
            \theta_{\mathrm{M},n} = 
            \frac{\mathds{1}^\top  \mat{C}^{-1}_{d_\mathrm{M}/r_\mathrm{d}}  \left. \vect{\frac{c}{H_0 \rd}}\right|_n}
            {\mathds{1}^\top  \mat{C}^{-1}_{d_\mathrm{M}/r_\mathrm{d}}  \mathds{1}}
        \end{equation}  is the weighted average over the values of $c/(H_0r_\mathrm{d})$ computed at each BAO redshift during iteration $n$ following eq.~\eqref{eq:c_H0rd_methodA}, and contained in the vector $\left. \vect{\frac{c}{H_0 \rd}}\right|_n$ of dimension $N_\mathrm{BAO}$, the number of BAO measurements. $\bm{z}_{\mathrm{BAO}}$ and $\bm{\frac{\dcom}{\rd}}$ are also vectors of dimension $N_\mathrm{BAO}$, and $\mat{C}_{\dcom /\rd}^{-1}$ is the (diagonal) inverse covariance matrix of the transverse BAO mode.

        \item The $\chi^2$ to both SNIa and the complete BAO data:
        \begin{subequations}
        \begin{align}
        \label{eq:chi2_tot}
            &\bm{\delta y_\mathrm{H}} = 
            \frac{\theta_{\mathrm{H},n}}{h_n}(\bm{z}_\mathrm{BAO}) - \bm{\frac{\mathrm{d_H}}{\mathrm{\rd}}},\\
            &\chi^2_{\mathrm{tot}} = \chi^2_\mathrm{SNIa} + \left( 
                                        \begin{array}{c}
                                        \displaystyle
                                        \bm{\delta y_\mathrm{M}} \\[10pt]
                                        \displaystyle
                                        \bm{\delta y_\mathrm{H}}
                                        \end{array}
                                        \right)^\top
                                        \mat{C}_{\mathrm{BAO}}^{-1}
                                    \left( 
                                        \begin{array}{c}
                                        \displaystyle
                                        \bm{\delta y_\mathrm{M}} \\[10pt]
                                        \displaystyle
                                        \bm{\delta y_\mathrm{H}}
                                        \end{array}
                                        \right),                    
        \end{align}
        \end{subequations}
        where $\mat{C}_{\mathrm{BAO}}^{-1}$ is the inverse covariance matrix of the BAO measurements, $\theta_{\mathrm{H},n}$ is computed similarly to eq.~\eqref{eq:theta_Mn}.
        However, a collection of reconstruction $\{ h_n(z) \}$ can only be defined through eq.~\eqref{eq:general_dimensionless_comoving_distance}, which requires a choice for the value of $\Omk[0]$ . We choose $ \Omk[0] = 0$ and can now use eq.~\eqref{eq:flat_dimensionless_comoving_distance} to calculate  $h_n(z)$ and eq.~\eqref{eq:c_H0rd_methodB_flat} for $\left. \vect{\frac{c}{H_0/\rd}}\right|_n$. In other words, $\chi^2_\mathrm{tot}$ can not be computed without resorting to a choice on the value of $\Omk[0]$. For this reason, as to not introduce bias in the results, we choose to limit this analysis to the two previous selection criterion. Results derived from this selection criterion could only be used as a consistency check.
        
    \end{enumerate}

    From there, we define :
    \begin{equation}
        \label{eq:delta_chi2}
        \Delta \chi^2_{\mathrm{data}} = \chi^2_{\mathrm{data}} - \chi^2_{\mathrm{data, f}\Lambda\mathrm{CDM}}      
    \end{equation}
    where the data considered for the fit are one of the three configurations described above (eq.~\ref{eq:chi2_SNIa}, \ref{eq:chi2_SNIa_dm} and \ref{eq:chi2_tot}) and $\chi^2_{\mathrm{data, f}\Lambda\mathrm{CDM}}$ is the $\chi^2$ of the flat $\Lambda$CDM model that best fits these data. Since $H_0$ is degenerated with $M_\mathrm{B}$ in the SNIa likelihood, and with $\rd$ in the BAO likelihood, we fix $H_0 = \SI{67.66} {km.s^{-1}.Mpc^{-1}}$ (taken from Planck 2018 \cite{Planck18}) and minimize the functions $\chi^2_{\mathrm{data, f}\Lambda\mathrm{CDM}}(\Omm[0], M_\mathrm{B})$ (for SNIa data) and $\chi^2_{\mathrm{data, f}\Lambda\mathrm{CDM}}(\Omm[0], r_\mathrm{d})$ (for BAO data).
    In the following, we apply the selection criteria $\Delta \chi^2_{\mathrm{data}} < 0$, so that we only use the reconstructions with a better fit to the data than the best fitting flat $\Lambda$CDM. Here, the flat $\Lambda$CDM best-fit act as an upper bound on the fit of the selected reconstructions, since reconstructions providing a poorer fit then the current concordance model are of little interest.

    Each selection criteria carries different information about the data. The first criterion, as originally used in \cite{ben2025}, only informs us about the fit of the reconstructions to the SNIa data. However, since the $\Ok$ diagnostic combines SNIa with BAO data, we introduce the second selection criterion to take into account the goodness of fit of the reconstructions to the transverse BAO mode and test the consistency between the two datasets.

\section{Results}
\label{sec:results}
    \subsection{Smoothing the SNIa}
    \label{sec:results_smoothing}

    \begin{figure}
        \centering
        \includegraphics[scale=1]{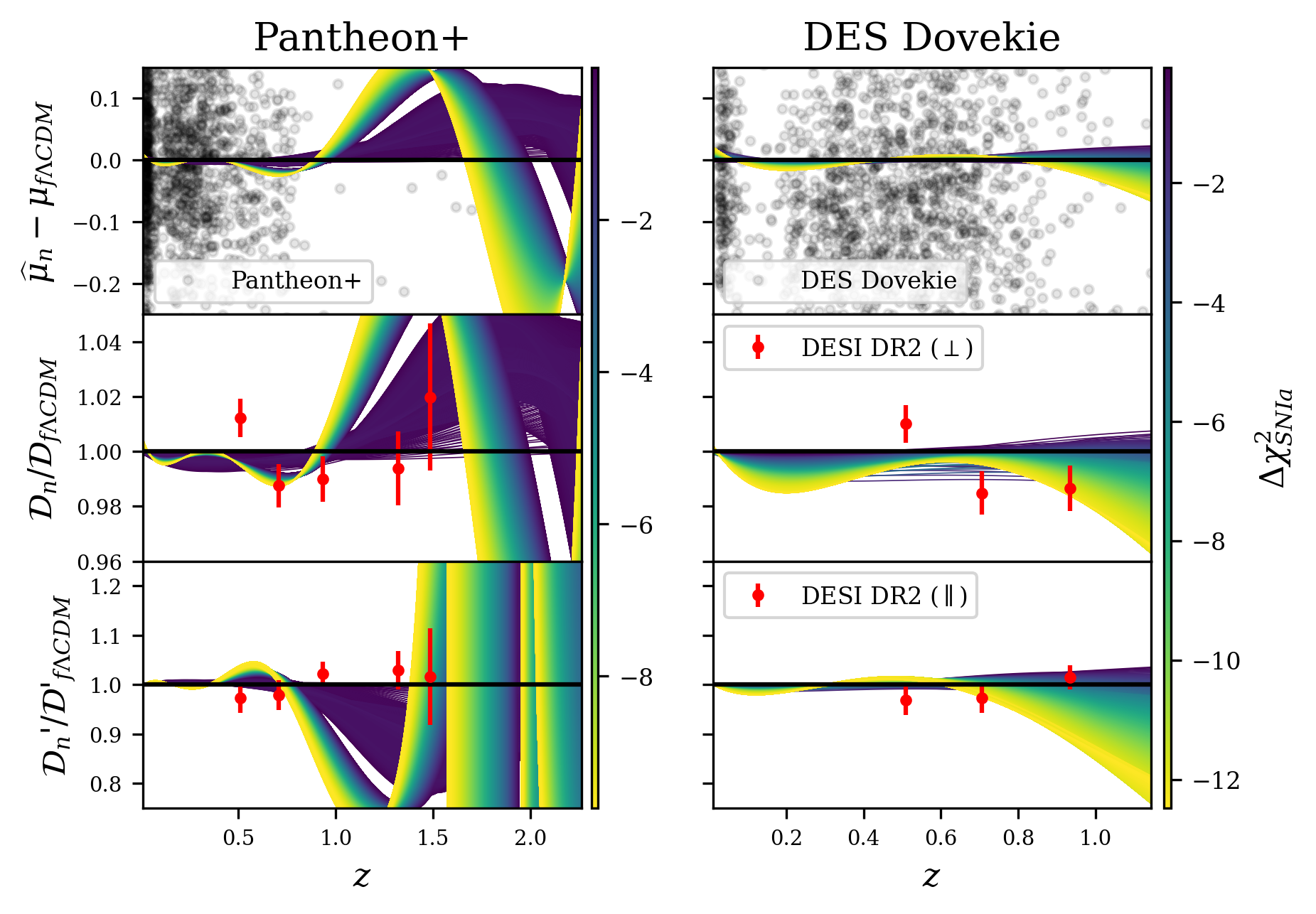}
        \caption{Iterative smoothing on SNIa data. The results for the Pantheon+ compilation are displayed in the left column, while the results for DES~Dovekie are displayed in the right column. The selection criterion is $\Delta \chi^2_{\mathrm{SNIa}} < 0$, and the reconstructions are colour-coded by $\Delta \chi^2_{\mathrm{SNIa}}$. The \textbf{top row} shows the results in the residual space $\mu_n(z) - \mu_{\mathrm{f \Lambda CDM}}(z)$, the \textbf{middle row} shows the dimensionless comoving distance $\mathcal{D}_n(z)$ of eq.~\eqref{eq:comoving_dist_to_distmod} normalised by $\mathcal{D}_{\mathrm{f \Lambda CDM}}(z)$, and the \textbf{bottom row} displays the derivative of dimensionless comoving distance with respect to redshift $\mathcal{D}_n'(z)$, normalised by $\mathcal{D}_{\mathrm{f \Lambda CDM}}(z)$ . The subscript $\mathrm{f\Lambda CDM}$ indicates a quantity derived from the flat $\Lambda$CDM model that best fits the data involved in the selection criteria. For illustration, the BAO data $d_\mathrm{M}/r_\mathrm{d}$ and $d_\mathrm{H}/r_\mathrm{d}$ are plotted in the middle and bottom row respectively. We use eq.~\eqref{eq:c_H0rd_methodA} and \eqref{eq:c_H0rd_methodB_flat}, and take the necessary value of $c/(H_0r\mathrm{d})$ from Planck 2018. In the case of the BAO the ratio $\mathcal{D}'(z)/\mathcal{D}'_{\mathrm{f \Lambda CDM}}(z)$ can be interpreted as $h_{\mathrm{f \Lambda CDM}}(z)/h(z)$.}    
        \label{fig:smoothing_SNIa}
    \end{figure}

    We perform the iterative smoothing algorithm on both Pantheon+ and DES~Dovekie dataset. 
    Various flat $\Lambda$CDM cosmologies are used as initial guesses.
    The upper row of Figure~\ref{fig:smoothing_SNIa} shows the results of the iterative smoothing of the distance modulus for the selection criterion $\Delta\chi^2_\mathrm{SNIa}<0$. 
    The selected reconstructions are colour-coded by $\Delta \chi^2_{\mathrm{SNIa}}$. 
    The left column shows the results of the iterative smoothing on Pantheon+ SNIa, while the right column shows the results for DES~Dovekie. The results are plotted in the residual space $\mu(z) - \mu_{\mathrm{f}\Lambda\mathrm{CDM}}(z)$, where $\mu_{\mathrm{f}\Lambda\mathrm{CDM}}$ is the distance modulus of the flat $\Lambda$CDM cosmology that best fits the data. 
    In the second row are the reconstructions of the dimensionless luminosity distance $\mathcal{D}_n$ normalized by the best fit $\mathcal{D}_{\mathrm{f}\Lambda\mathrm{CDM}}$, using eq.~\eqref{eq:comoving_dist_to_distmod}. 
    Finally, in the third row, we display the reconstructions of $\mathcal{D}_n'$ normalised by the best fit derivative $\mathcal{D}'_{\mathrm{f}\Lambda\mathrm{CDM}}$. When $\Omk[0]=0$, this quantity is equal to the inverse of normalized Hubble parameter $h_{\mathrm{f}\Lambda\mathrm{CDM}}/h_n$ (see eq.~\ref{eq:flat_dimensionless_comoving_distance}). For information purposes, we over-plotted the BAO data from DESI~DR2 in the two bottom rows of the figures using eq \eqref{eq:c_H0rd_methodA} and \eqref{eq:c_H0rd_methodB_flat} to convert the two BAO modes into $\mathcal{D}$ and $\mathcal{D}'$ respectively. The $c/(H_0\rd)$ value is taken from Planck 2018 \cite{Planck18}. We emphasize that these points are not included in the smoothing procedure, and are here only as illustration.
    The Pantheon+ reconstructions that best fit the SNIa data (in yellow) appear to be those that fit the less the BAO data, plotted in red in the two bottom rows. 
    At high redshift, where the data become sparse, the reconstructions strongly deviate from the best-fitting flat $\Lambda$CDM.

    \begin{figure}
        \centering
        \includegraphics[scale=1]{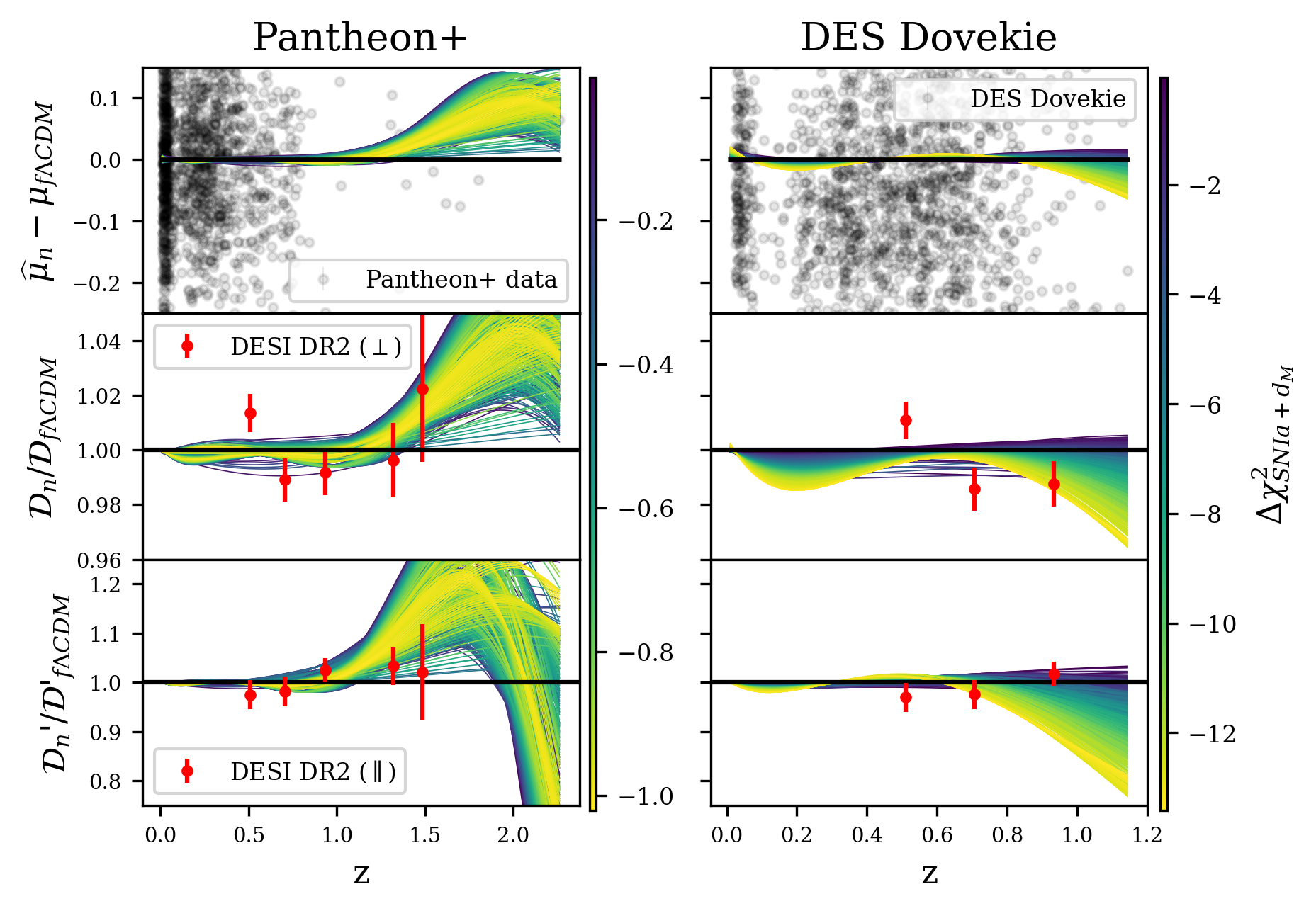}
        \caption{Same as Fig. \ref{fig:smoothing_SNIa}, but using selection criterion $\Delta \chi^2_{\mathrm{SNIa} + \dcom} < 0$}
        \label{fig:smoothing_dm}
    \end{figure}

    \begin{figure}
        \centering
        \includegraphics[scale=1]{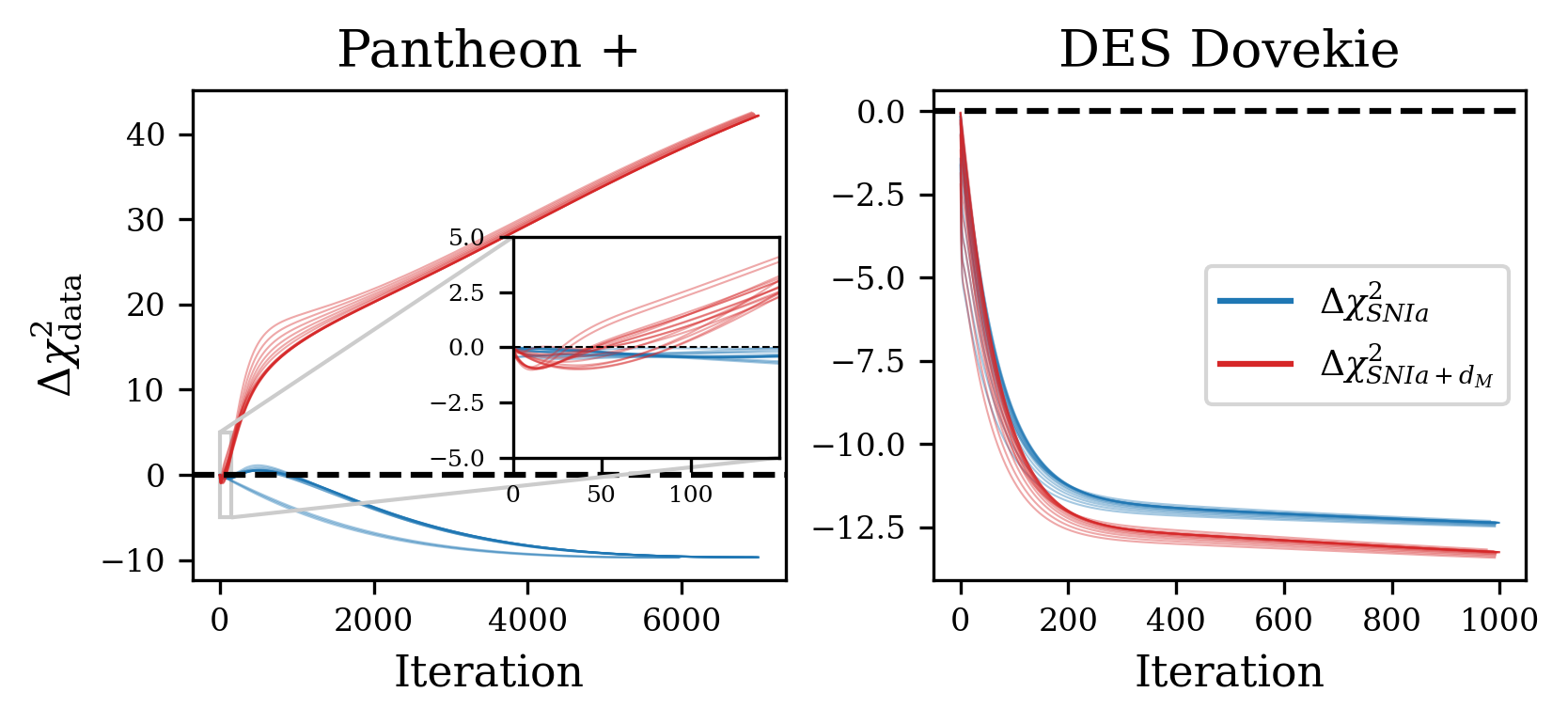}
        \caption{Iterative evolution of $\Delta \chi^2_{\mathrm{data}}$. The blue curves are the goodness of fit of the reconstruction to the SNIa data alone, the red curves to both SNIa and the transverse BAO mode, and the green curves to both the SNIa and the complete BAO data. The various curves of same colour correspond to different initial guesses. The black dotted horizontal line represents the selection criterion. Reconstruction crossing into $\Delta \chi^2_{\mathrm{data}} > 0$ are not be kept for the $\Ok$ diagnostic.} 
        \label{fig:chi2_comparison_desd}
    \end{figure}

    When changing the selection criterion to $\Delta \chi^2_{\mathrm{SNIa} + \dcom} < 0$ in the left column of Fig. \ref{fig:smoothing_dm}, we can see, as expected, that the reconstructions that best fit the SNIa data in Fig. \ref{fig:smoothing_SNIa} are left out. In the second row of the left column, we can see that the remaining reconstructions have a good fit to the transverse BAO data point.  
    Adding information from the transverse BAO mode strongly affect the Pantheon+ collection of selected reconstructions. 
    This can be seen in the left panel of Fig. \ref{fig:chi2_comparison_desd}, which shows the evolution of $\Delta\chi^2_\mathrm{data}$ for the three considered selection criteria.  
    Each solid line corresponds to one choice of initial condition.  
    The blue solid lines show the evolution of $\Delta\chi^2_{\mathrm{SNIa}}$. 
    Similarly, the red lines correspond to the evolution of $\Delta \chi^2_{\mathrm{SNIa} + \dcom}$. 
    As expected, $\Delta\chi^2_\mathrm{SNIa}$ (in blue) becomes more and more negative as the iterative smoothing progresses and the smooth reconstructions get closer to the data for both Pantheon+ and DES~Dovekie.
    In the case of DES~Dovekie, the reconstructed comoving distances are also in good agreement with the transverse BAO mode, leading to a slightly lower $\Delta\chi^2_{\mathrm{SNIa} + \dcom}$, and $\Delta\chi^2_{\text{SNIa}+\dcom}\simeq \Delta\chi^2_\text{SNIa}$.
    
    However, the disagreement between the reconstructed distances from Pantheon+ and the comoving distances from DESI~DR2 can be seen in the red lines of the left panel. 
    In this case, $\Delta\chi^2_{\mathrm{SNIa} + \dcom}$ first decreases during the first few iterations, and then starts increasing and diverges away from zero. 
        This reveals a tension between the Pantheon+ and the DESI~DR2 data : the Pantheon+ data, which become sparse at redshift $z>1$, drive the reconstructions away from the BAO data. 
    This is not the case for the DES~Dovekie collection: even if $\Delta \chi^2_{\mathrm{SNIa} + \dcom}>\Delta \chi^2_{\mathrm{SNIa}}$, $\Delta \chi^2_{\mathrm{SNIa} + \dcom}$ does not cross into positive values. 
    Consequently, the DES~Dovekie collection remains virtually unchanged under the new selection criterion.

    \subsection{Curvature test}
    \label{sec:results_curvature_test}

    \begin{figure}
        \centering
        \includegraphics[scale=1]{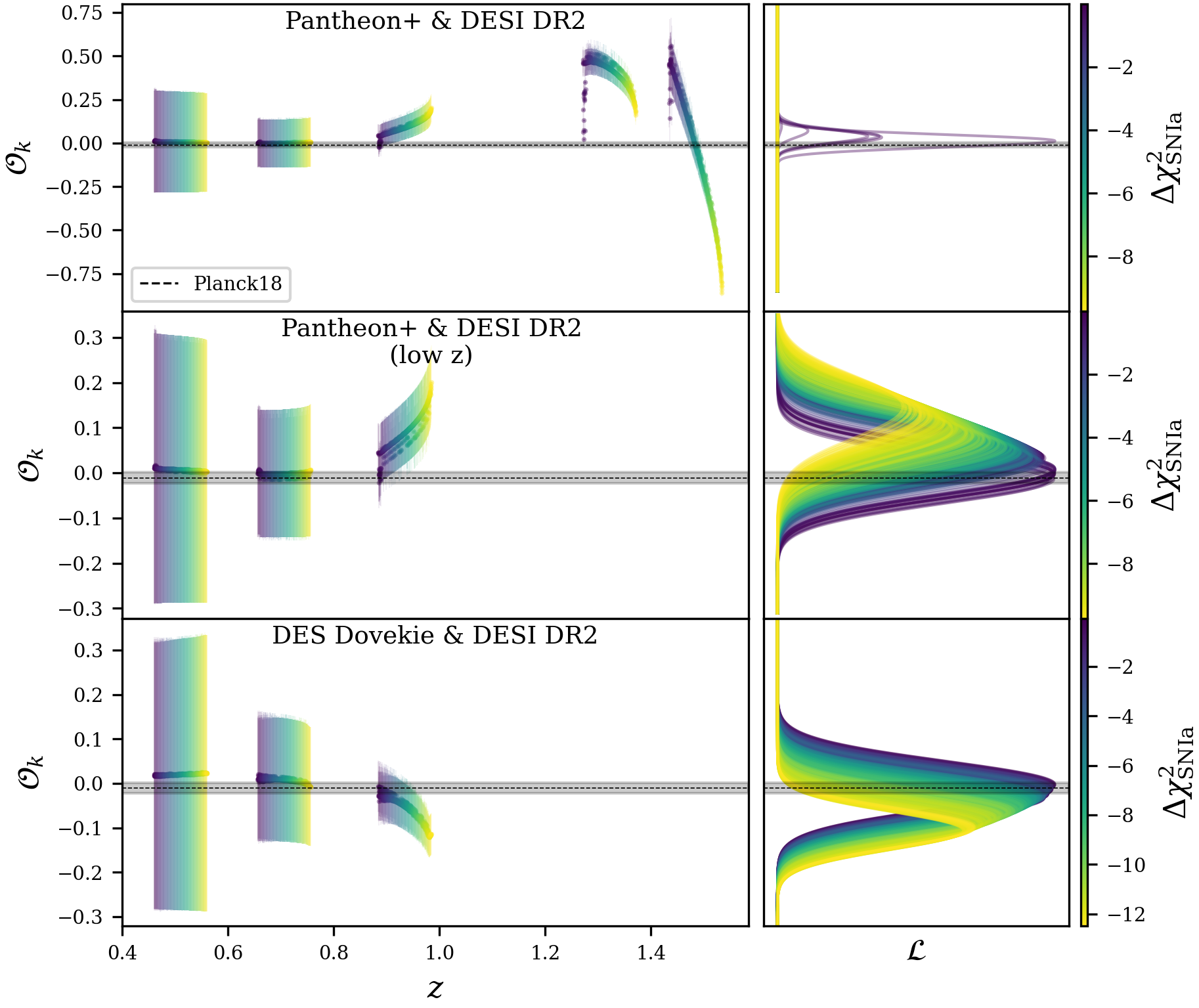}
        \caption{Results for the $\Ok$ diagnostic, with selection criterion $\Delta \chi^2_{\mathrm{SNIa}} < 0$. From top to bottom are displayed the results for the combination of Pantheon+ \& DESI~DR2, Pantheon+ \& DESI~DR2 truncated at $z = 1.13$  and DES~Dovekie \& DESI~DR2. 
        \textbf{Left column:} $\Ok$ diagnostic of each selected reconstructions at the redshift of the DESI~DR2 BAO, along with their $1\sigma$ errors and colour-coded by $\Delta \chi^2_{\mathrm{SNIa}}$. For readability purpose, the result of each reconstruction is plotted with a slight, distinct offset in $z$ around the original BAO redshift.
        \textbf{Right column:} likelihoods of each reconstruction, colour-coded by $\Delta \chi^2_{\mathrm{SNIa}}$.}
        \label{fig:Ok_SNIa}
    \end{figure}

    The results for the reconstruction of the $\Ok$ diagnostic using the reconstructions of Fig. \ref{fig:smoothing_SNIa}, are displayed in Fig. \ref{fig:Ok_SNIa}, where the selection criterion applied is $\Delta \chi^2_{\mathrm{SNIa}} < 0$. At each BAO redshift $z_i  =( \vect{z}_\mathrm{BAO})_i$, we obtain $N_\mathrm{rec}$ values of $\Ok(z_i)$, where $N_\mathrm{rec}$ is the total number of reconstructions that passed the selection criterion. In order not to overload the plot and help readability, only 500 reconstructions were plotted among the $N_\mathrm{rec}$ selected reconstructions. To ensure that the plotted sample represents the diversity of results, the 500 plotted reconstructions were sampled uniformly in $\Delta \chi^2_{\mathrm{SNIa}}$. 
    The left column shows the $\Ok$ diagnostic for each reconstruction, with its 1$\sigma$ uncertainty, propagated as indicated in \cite{ben2025}. 
    The error bars are larger at lower redshift due to the division by $\mathcal{D}_n^2(z)$ during error propagation. The results were computed at the BAO redshifts. For readability, and given the high number of reconstructions to plot, we apply a small, unique offsets in $z$ to each reconstruction, so that the corresponding estimates of $\Ok(z)$ are shown side-by-side rather than overlapping.
    The right column shows the likelihoods $\mathcal{L} = \exp{(-\chi^2_{\Ok}/2)}$ of each reconstruction, where $\chi^2_{\Ok}$ is the $\chi^2$ of a constant to the $\Ok$ diagnostic. As previously, the reconstructions are colour-coded by $\Delta \chi^2_{\mathrm{SNIa}}$. The first row displays the results for the combination of Pantheon+ \& DESI~DR2. In the middle row, we restrict the redshift range to the 3 lowest BAO redshifts. Finally, the results for DESY~Y5 \& DESI DR2 are displayed in the bottom row.

    In the Pantheon+ \& DESI~DR2 case, the reconstructions with more negative $\Delta \chi^2_\mathrm{SNIa}$ (in yellow), which are the most favoured by the SNIa data, appear to deviate significantly from constant, and thus from FLRW, at high redshift. Their corresponding likelihoods have indeed a very low peak, especially when compared to the likelihoods of reconstructions with less negative $\Delta\chi^2_\mathrm{SNIa}$ (in violet). The latter reconstructions, having a $\Delta \chi^2_\mathrm{SNIa}$ close to zero, provide a comparatively poorer fit to the data, but their high-peaked likelihood indicate a far better consistency with the FLRW metric.
    When reconstructions are inconsistent with the FLRW metric, the $\Ok$ diagnostic cannot be extended as a test on curvature.
    Consequently, we decide to filter out the reconstructions that are inconsistent with FLRW before moving into the curvature test.
    To do so, we perform a $p$-value test on the collection of reconstructions. 
    We assume that $\chi^2_{\Ok}$ follows a $\chi^2_{\nu}$ distribution of degree of freedom $\nu = N_{\mathrm{BAO}}-p$, where $N_{\mathrm{BAO}}$ is the number of BAO data points and $p=1$ is the number of parameters in the model (here, the model is a constant fitting). 
    We choose $\alpha = 0.05$, and reject all reconstructions with $ p < \alpha $, so that we can affirm with 95\% confidence that the reconstructions that pass the test are consistent with a constant $\Ok$, and that this $\Ok$ can be interpreted as an estimation of \Omk[0]. 
    In Table \ref{tab:recap_all} we report ${\Omega}_{k,0}^\text{med}$ the median of these constants \Omk[0], with the spread of all FLRW-consistent reconstructions $\Delta \Omk[0] = \max \Omk[0]-\min\Omk[0]$, and the median $1\sigma$ errors ${\sigma}_{\Omk[0]^\text{med}}$, as well as the median value of $c/(H_0r_\mathrm{d})$ computed with eq.~\eqref{eq:c_H0rd_methodA}. Our various estimations of $c/H_0r_\mathrm{d}$ are all consistent within 2$\sigma$ with the value from Planck 2018 $c/H_0r_\mathrm{d} = 30.26 \pm 0.28$~\cite{Planck18}.

    Only $\sim$1\% of the Pantheon+ \& DESI~DR2 reconstructions pass the $p$-value test. The remaining reconstructions are consistent with the FLRW metric, with ${\Omega}_{k,0}^\text{med} = 0.064^{+0.048}_{-0.099} \pm 0.038$, where the first uncertainty is the spread of the central value of ${\Omega}_{k,0}$ over all the reconstructions, and the second uncertainty is the median of the 1$\sigma$ uncertainties (see Table~\ref{tab:recap_all}). The maximum deviation from flatness is 2.76$\sigma$, while maintaining a slight preference for positive \Omk[0]. When considering only the lower redshift range ($z<1.13$) for the BAO data to protect the $\Ok$ diagnostic from edge effects and the sparsity of high-$z$ Pantheon+ data, 100\% of the reconstructions are consistent with FLRW, supporting the conclusion that the deviation from FLRW in the Pantheon+ \& DESI~DR2 data combination is driven by the high redshift supernovae. The reconstructions best-fitting the SNIa still present a lower likelihood, and they drive $\Omk[0]$ away from flatness and toward positive values. 
    The median value is $\Omk[0]^\text{med} = 0.102^{+0.056}_{-0.129}\pm 0.064$, with a maximum deviation from flatness of 2.28$\sigma$.

\begin{figure}
    \centering
    \includegraphics[scale=1]{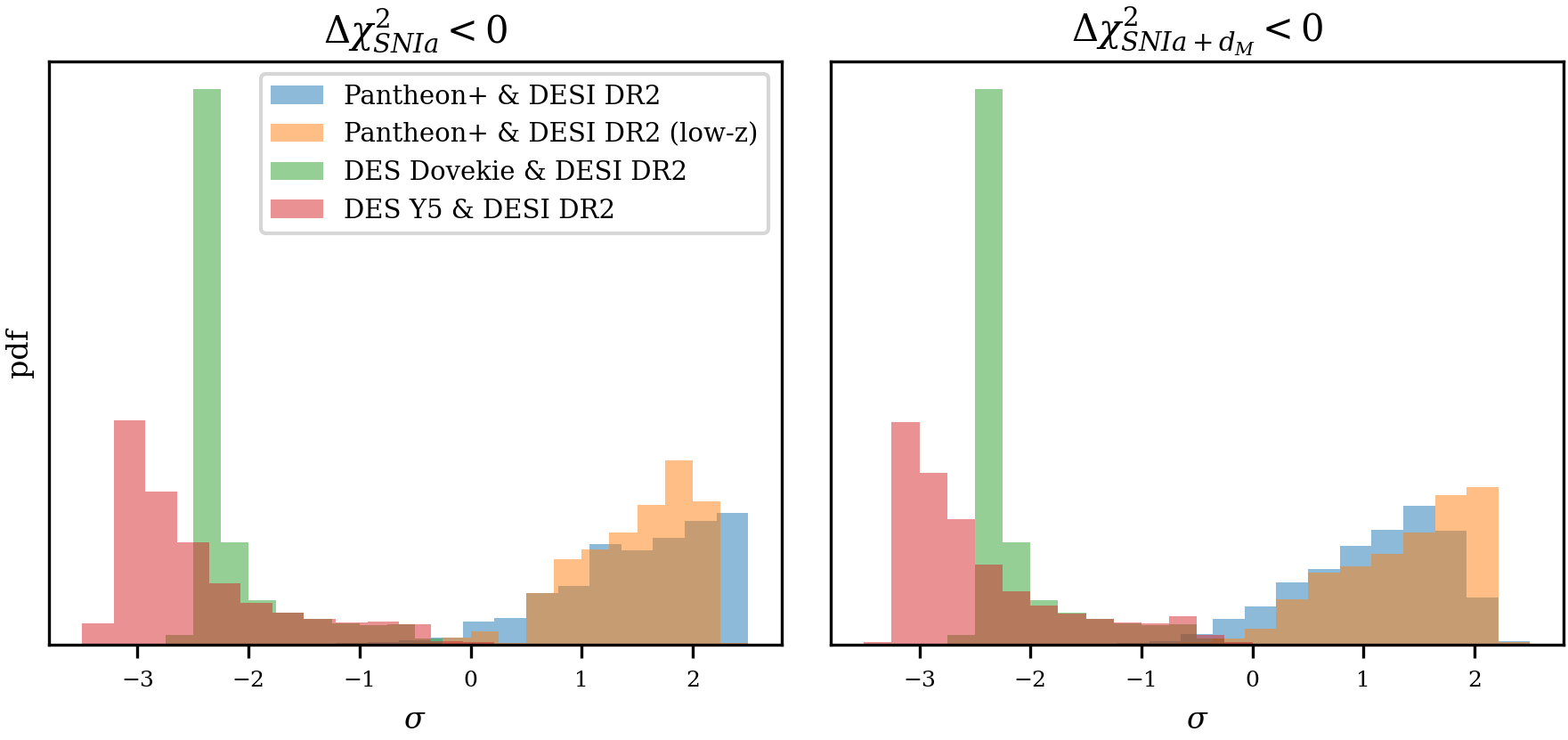}
    \caption{For each selection criterion, histogram of the deviation from flatness of the reconstructions that passed the $p$-value test and are therefore consistent with the FLRW metric.}
    \label{fig:deviation}
\end{figure}

    On the contrary, DES~Dovekie \& DESI~DR2 have an excellent consistency with the FLRW metric. Indeed, this combination of data results in 100\% of the reconstructions passing the $p$-value test. The median constant is ${\Omega}_{k,0}^\text{med} = -0.102^{+0.100}_{-0.005}\pm 0.043$ with a maximum deviation from flatness of -2.51$\sigma$, which correspond to reconstructions with the best fit to the SNIa data (in yellow).  We can see on the left panel of Fig.~\ref{fig:deviation} that all reconstructions remain consistent with flatness within 3$\sigma$. 
    Similarly to the Pantheon+ \& DESI~DR2 (low-$z$) results, these reconstructions best fitting the DES~Dovekie SNIa (in green) tend to have a lower likelihood and are centered around higher $\Ok$ than the reconstruction with a higher $\Delta \chi^2_{\mathrm{SNIa}+d_\mathrm{M}}$. 
    The median values ${\Omega}_{k,0}^\text{med}$ of the maximum likelihoods are recapitulated in Table~\ref{tab:recap_all} together with the median standard deviation ${\sigma}_{\Omk[0]^\text{med}}$ and the total spread $\Delta \Omk[0]$. 
    
    \begin{figure}
        \centering
        \includegraphics[scale=1]{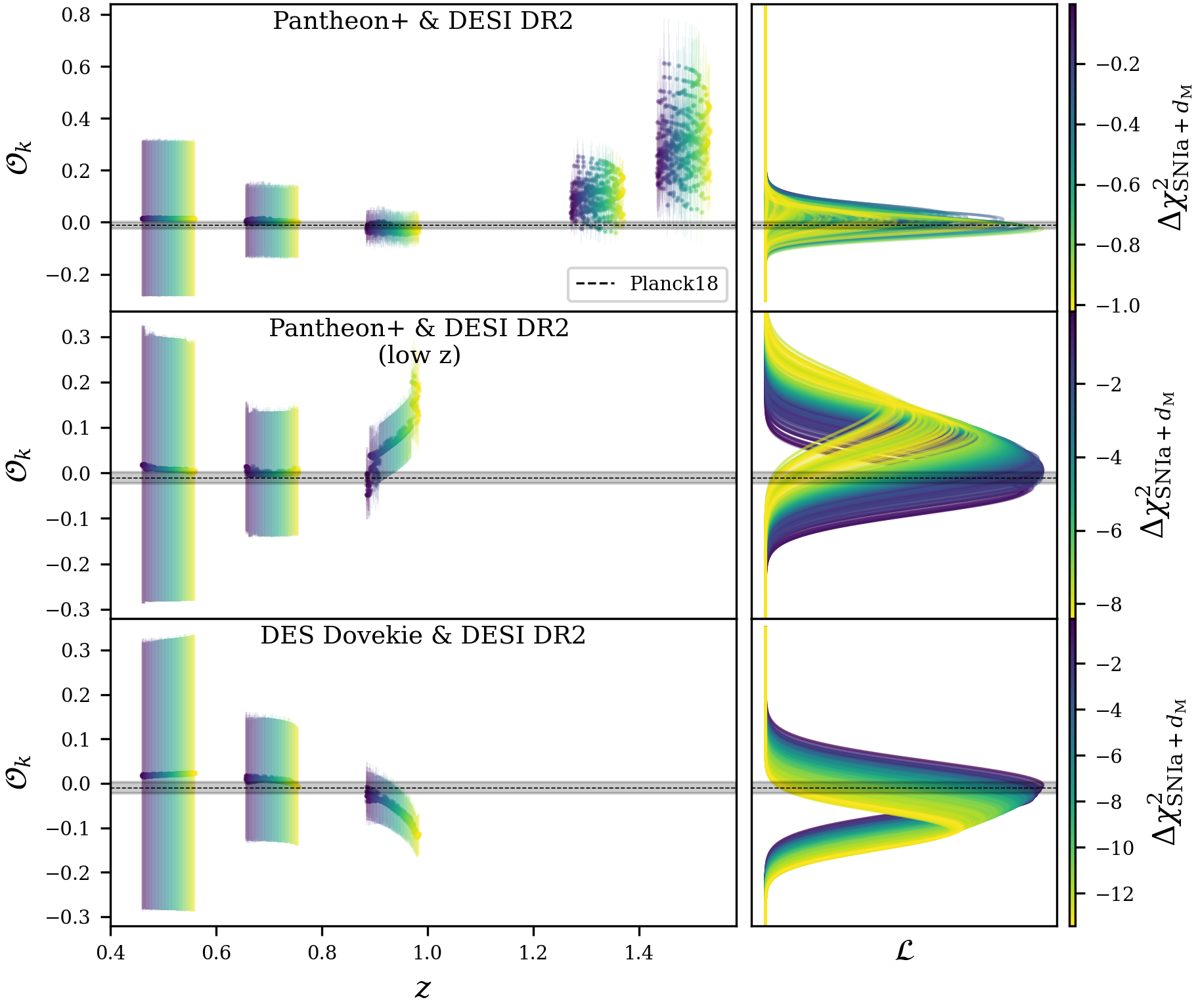}
        \caption{Results for the $\Ok$ diagnostic, with selection criterion is $\Delta \chi^2_{\mathrm{SNIa} + \dcom} < 0$. On the right column are the likelihoods of each reconstruction, colour-coded by $\Delta \chi^2_{\mathrm{SNIa} + \dcom}$.}
        \label{fig:Ok_dm}
    \end{figure}

    Changing the selection criterion to $\Delta \chi^2_{\mathrm{SNIa} + \dcom} < 0$, we display the resulting $\Ok$ diagnostic in Figure \ref{fig:Ok_dm}.  We saw in section \ref{sec:results_smoothing} that adding transverse BAO information to the $\chi^2$ selection leads to a significant reduction in the number of selected reconstructions for the combination of Pantheon+ \& DESI~DR2 data. The curved patterns of the $\Ok$ points at $z=1.32$ and $z=1.48$ in the Pantheon+ \& DESI DR2 case are artefacts of the evolution $\Delta \chi^2_{\mathrm{SNIa} + \dcom}$ with iterations (in red in the left panel of Fig. \ref{fig:chi2_comparison_desd}). 
    The decrease and further increase in $\Delta \chi^2_{\mathrm{SNIa} + \dcom}$ as the value of $\Ok$ comes closer to the one favoured by SNIa data create the curved shape of the data points, which are ordered from left to right by decreasing $\Delta \chi^2_{\mathrm{SNIa} + \dcom}$. Indeed, we can see that the reconstructions significantly deviating from constant in the top row of Fig. \ref{fig:Ok_SNIa} are absent from Fig. \ref{fig:Ok_dm}, and under selection criterion $\Delta \chi^2_{\mathrm{SNIa} + \dcom} < 0$, $\sim$82\% of the remaining reconstructions from Pantheon+ \& DESI~DR2 pass the $p$-value test, i.e. are consistent with the FLRW metric.
    The median value of the $\Ok$ diagnostic is ${\Omega}_{k,0}^\text{med} = 0.045^{+0.045}_{-0.081} \pm 0.038$ (see Table \ref{tab:recap_all}), and the highest deviation from flatness is 2.29$\sigma$. The preference for positive \Omk[0] remains.
    
    In Pantheon+ \& DESI~DR2 low-$z$ data, for which 100\% of reconstruction are consistent with FLRW, the median is ${\Omega}_{k,0}^\text{med} = 0.095^{+0.063}_{-0.136}  \pm 0.063$. In this case,  and unlike the $\Delta \chi^2_{\mathrm{SNIa}}$ selection, the calculation of $\Delta \chi^2_{\mathrm{SNIa} + \dcom}$ was carried over three BAO measurements instead of five. Although the high redshift DESI data have been truncated, the reconstructions have still been smoothed on the full redshift range over the Pantheon+ dataset, keeping the SNIa part of $\chi^2_{\mathrm{SNIa} + \dcom}$ unchanged. This restriction of the BAO data to the 3 lowest redshift bin ($z \lesssim 1$) affects the value of $\Delta \chi^2_{\mathrm{SNIa} + \dcom}$, and therefore the selection itself. Some of the Pantheon+ reconstructions that did not meet the selection criterion $\Delta \chi^2_{\mathrm{SNIa} + \dcom}<0$ when considering all 5 BAO points are accepted once the high redshift data are removed. These additional reconstructions drive the estimation of ${\Omega}_{k,0}^\text{med}$ toward a higher positive value. The highest deviation from flatness being 2.23$\sigma$.
    
    Unlike Pantheon+ \& DESI~DR2, DES~Dovekie \& DESI~DR2 are in better agreement with each other : the transition from the bottom row of Figure \ref{fig:Ok_SNIa} to the bottom row of \ref{fig:Ok_dm} does not significantly reduce the selection of reconstructions. We can see in Table \ref{tab:recap_all} that ${\Omega}_{k,0}^\text{med} = -0.102^{+0.099}_{-0.005}\pm 0.043$, remains virtually unaffected by the change of selection criterion, as well as the -3.28$\sigma$ maximum deviation from flatness. The conclusions for DES~Dovekie \& DESI~DR2 remain unchanged: FLRW is not ruled out, a negative $\Omk[0]$ is still preferred, and all reconstructions are consistent with $\Omk[0] = 0$ within 3$\sigma$ in absolute value (see middle panel of Fig. \ref{fig:deviation}).

\renewcommand{\arraystretch}{1.4}
\begin{table}
    \centering
    \begin{tabular}{c c c c}
    \toprule
    $\Delta\chi^2$ criterion 
    & Data combination 
    & ${\Omega}_{k,0}^\text{med}$ 
    & $\frac{c}{H_0 r_\mathrm{d}}$\\
    \midrule
    
    \multirow{3}{*}{$\Delta\chi^2_{\mathrm{SNIa}}<0$}
    & P+ \& DESI~DR2 
    & $0.064^{+0.048}_{-0.099} \pm 0.038$ & $29.937^{+0.173}_{-0.052} \pm 0.140$\\
    
    & P+ \& DESI~DR2 (low-$z$) 
    & $0.102^{+0.056}_{-0.129} \pm 0.064$ & $30.104^{+0.091}_{-0.87} \pm 0.158$ \\
    
    & DES~Dov. \& DESI~DR2 
    & $-0.102^{+0.100}_{-0.005} \pm 0.043$ & $30.254^{+0.014}_{-0.422} \pm 0.159$\\
    \midrule
    
    \multirow{3}{*}{$\Delta\chi^2_{\mathrm{SNIa}+\dcom}<0$}
    & P+ \& DESI~DR2 
    & $0.045^{+0.045}_{-0.081} \pm 0.038$ & $30.051^{+0.277}_{-0.103} \pm 0.141$ \\
    
    & P+ \& DESI~DR2 (low-$z$) 
    & $0.095^{+0.063}_{-0.136} \pm 0.063$ & $30.112^{+0.312}_{-0.280} \pm 0.158$ \\
    
    & DES~Dov. \& DESI~DR2 
    & $-0.102^{+0.099}_{-0.005} \pm 0.043$ & $30.254^{+0.014}_{-0.380} \pm 0.159$\\
    
    
    
    \bottomrule
    \end{tabular}
    
    \caption{Recap table of ${\Omega}_{k,0}^\text{med}$, along with the spread of all reconstructions and the median $1\sigma$ error, and of median value of $c/(H_0r_\mathrm{d})$ with $1\sigma$ error, for each data combination and each $\Delta\chi^2$ selection criterion. Results were calculated using only the reconstructions consistent with the FLRW metric with 95\% confidence.}
    \label{tab:recap_all}
\end{table}

\section{Discussion and conclusion}
\label{sec:conclu}

We performed a model-independent litmus test of the FLRW metric and the flatness of the universe. Using an iterative smoothing algorithm on Pantheon+ and DES~Dovekie data, we reconstructed $\mathcal{D}_n(z)$ and $\mathcal{D}_n'(z)$. Then we evaluated these quantities at the redshifts of the BAO data from DESI~DR2 and combined them with the transverse and radial modes $d_M/\rd$ and $d_{\mathrm{H}}/\rd$ to obtain the $\mathcal{O}_k$ diagnostic (eq.~\ref{eq:Ok}~\& \ref{eq:Theta_combined}), a litmus test for both the FLRW metric and  flatness. 

For the Pantheon+ \& DESI~DR2 data combination, the reconstructed $\Ok$ diagnostics, because they present a deviation from constant at high-redshift, are for the most part inconsistent with the FLRW metric. 
The closer the reconstructions are to the (SN) data, the more important this deviation is. Only $\sim$1\% of reconstructions do not rule out the FLRW metric, and these few reconstructions are also consistent with flatness within 3$\sigma$, with a preference for positive $\Omk[0]$.
However, when considering the fit to the transverse mode of the BAO $d_\mathrm{M}/\rd$, the reconstructions excluding the FLRW metric are rejected by the selection criterion $\Delta \chi^2_{\mathrm{SNIa} + \dcom}<0$, showing an inconsistency between the Pantheon+ SNIa and the DESI~DR2 BAO. 
Under this new criterion, 100\% of the selected reconstructions are consistent with FLRW, and are also consistent with flatness within 3$\sigma$.
When restricting the redshift range to $z<1.13$ (i.e. the BAO datapoints with the two highest redshifts are removed from the DESI DR2 data), the reconstructed distances from SNIa and the BAO transverse distances become fully consistent with an FLRW universe, supporting that the tension with the FLRW metric is driven by the high-redshift end of the Pantheon+ SNIa, where the data are sparse and less reliable.  Supernovae with $z>1$ have typically larger systematic uncertainties due to selection effects such as the Malmquist bias. 
A study of the Pantheon dataset \cite{ben_linder2019} showed that a simple linear correction for Malmquist bias efficiently reduces deviations in the high-redshift reconstructions, although this correction is not statistically required by the data. Similarly, the high-redshift BAO data points removed in this part of the analysis are those most affected by modelling and systematic uncertainties \cite{Abdul_Karim_2025}.

The DES~Dovekie SNIa data do not show such tension with the DESI~DR2 BAO data. The selection of reconstructions is barely affected by the change in the selection criterion from $\Delta \chi^2_\mathrm{SNIa} < 0$ to $\Delta \chi^2_{\mathrm{SNIa} + \dcom}<0$, indicating good agreement between the two data sets. 
In both cases, the reconstructed $\Ok $ diagnostic does not rule out the FLRW metric, and is consistent with flatness within 3$\sigma$.

Our results are consistent with the Planck 2018 \cite{Planck18} results, and with previous model-independent tests of the FLRW metric involving SNIa and BAO data, which are in agreement with FLRW and flatness. 
The $\Ok$ diagnostic conducted in \cite{Dias_2025} with Pantheon+ data and cosmic chronometers found no evidence for a non-FLRW, non-flat universe.
Slight preference for various sign of $\Omk[0]$ appears depending on the data combination and the method, but these preference are never statistically significant. 
In \cite{Montanari_2017}, a negative $\Omk[0]$ is slightly favoured, while in \cite{Collet_2019} (Strong lensing sources and the Pantheon compilation), positive $\Omk[0]$ is favoured ($\Omk[0] = 0.12^{+0.27}_{-0.25}$). 
Similarly, null-test of the FLRW metric with Pantheon+ SNIa and DESI DR2 BAO found no deviation from FLRW, and no strong evidence for curvature within the FLRW metric \cite{Dinda_2025}. 
As for tests of spatial curvature only, they usually find no significant deviation from flatness. Pantheon+ SNIa combined to Cosmic Chronometers (CCH) in \cite{Liu_2020} give $\Omk[0] = -0.02 \pm 0.14$, and  in \cite{Favale_2025} DES~Dovekie \& DESI~DR2  combined with CCH resulted in $\Omk[0] = -0.143 \pm 0.085 $ ($1.7\sigma$ deviation from flatness) and $\Omk[0] = -0.107^{+0.079}_{-0.085}$ when replacing the DES~Dovekie data by Pantheon+, and combined to Fast Radio Bursts in \cite{Fortunato_2025} the DESI DR2 data indicate $\Omk[0] = -0.09^{+0.24}_{-0.31}$. The model independent measurement of $\Omk[0]$ in \cite{Du_2025} combines DESI DR2 with DES~Y5 SNIa, strong gravitational lensing, cosmic chronometers and gamma ray bursts, and finds $\Omk[0]=0.001\pm0.038$. Our estimations of $c/H_0r_\mathrm{d}$ are also consistent with $\frac{c}{H_0r_\mathrm{d}}=29.90\pm0.33$ from \cite{Jiang_2024}, a non-parametric inference of the constant using DESI~DR2 and Pantheon+ data.

Since high redshift Pantheon+ SNIa data revealed themselves to be quite unreliable for model-independent reconstructions of the expansion history, the robustness of the $\Ok$ diagnostic is confined to the low-redshift regime. To extend this model independent test of the FLRW metric and spatial curvature, we will require better high SNIa observations, or alternative cosmological probes with reliable data above $z \gtrsim 1$. Standard sirens are promising candidates for future tests, especially the upcoming third generation of ground based detector \cite{Cao_2019}. Generated mock data of upcoming Gravitational Waves survey going up to redshift $z\sim5$ result in a significant improvement of the constraints on cosmological parameters \cite{GW_mock}.  If these improvements remain highly model-dependent, standard sirens can also meet, and even exceed the constraint on $\Omk[0]$ from DES in a model independent framework \cite{Liao_2019}.

It is important to note that the results of this work depend on the choice of the smoothing scale $\Delta = 0.3$. Alternative results of the analysis with $\Delta = 0.2$ and $\Delta=0.4$ are available in Appendix~\ref{sec:smoothing_scale}. A lower smoothing scale tends to fit the data closely and reinforces the preferences of the datasets for positive or negative values of $\Omk[0]$. On the contrary, a higher smoothing scale ($\Delta =0.4$) results in reconstruction less affected by the particularities of each dataset, and the resulting $\Ok$ diagnostic present a stronger agreement both with flatness and other datasets. The value $\Delta=0.3$ was used in various studies involving different SNIa compilations such as the Joint Light-curve Analysis \cite{ben_arman2017} (740 SNIa), the Pantheon sample \cite{arman_ben2018} (1048 SNIa) or mock LSST data \cite{ben2025} (14 bins). Since the optimal value of $\Delta$ depends on the size and the quality of the sample, it seems relevant to re-evaluate the optimal value of $\Delta$ specifically for the Pantheon+ and DES~Dovekie compilations in future studies. Alternatively, we could also implement a dynamical smoothing scale $\Delta(z)$, as has already been suggested in \cite{arman2006}. The latter option seems to be particularly interesting given the important impact that the sparsity of the high-$z$ Pantheon+ data appears to have on our results.

\appendix

\setcounter{figure}{0}
\renewcommand{\thefigure}{\thesection.\arabic{figure}}
\setcounter{table}{0}
\renewcommand{\thetable}{\thesection.\arabic{table}}

\section{Normalisation method}
\label{sec:normalisation_method}

    Pantheon+ and DES~Dovekie provide respectively $m_\mathrm{B}$ and $\mu$, the iterative smoothing algorithm reconstruct directly these quantities.
     
     Consequently, the iterative smoothing algorithm, when applied to the Pantheon+ compilation, returns a collection of reconstructions $\{\widehat{m_\mathrm{B}}_{,n}(z)\}$. To obtain a collection of reconstructed $\{\widehat{\mu}_n(z)\}$, a value must be assumed for $M_\mathrm{B}$, the absolute magnitude of SNIa. As a result, a Pantheon+ reconstruction $\widehat{\mu}_n(z) = \widehat{m_\mathrm{B}}_{,n}(z) - M_\mathrm{B} $ is reconstructed up to an additive constant. Mathematically, 
     \begin{equation}
        \label{eq:true_dist_mod}
         \mu_n^{\mathrm{true}}(z) = \widehat{\mu}_n(z) + \Delta M_\mathrm{B},
     \end{equation}
     where $\Delta M_\mathrm{B} = M_\mathrm{B}^\mathrm{true} - M_\mathrm{B}$

     Eq.~\eqref{eq:true_dist_mod} also holds true for the DES~Dovekie collection of reconstructions. Indeed, the value of $M_\mathrm{B}$ chosen for the compilation might be off the true value by $\Delta M_\mathrm{B} = M_\mathrm{B}^\mathrm{true} - M_\mathrm{B}$.

     In theory, following eq~\eqref{eq:comoving_dist_to_distmod}:
    \begin{equation}
    \label{eq:D_from_reconstruction}
    \mathcal{D}_n(z) 
    = \frac{H_0}{c} \, \frac{10^{\frac{\mu_n^\mathrm{true}}{5} - 5}}{1+z} 
    = \frac{H_0}{c} 10^{\frac{\Delta M_\mathrm{B}}{5}}
    \frac{10^{\frac{\hat{\mu}_n}{5} - 5}}{1+z}
    = \frac{H_0}{c} \, 10^{\frac{\Delta M_\mathrm{B}}{5}} \, \mathcal{\widehat{D}}_n(z).
    \end{equation}
    Substituting with our reconstructions, we see that the additive constant propagate into $\mathcal{D}_n$ as a multiplicative constant, which we will denote as $A \equiv \frac{H_0}{c} \, 10^{\frac{\Delta M_B}{5}}$, such as:
    \begin{align}
        \label{eq:collection_D}
        \mathcal{D}_n(z) &= A \, \mathcal{\widehat{D}}_n(z) \\
        \label{eq:collection_Dprime}
        \mathcal{D}_n'(z) &= A \,  \mathcal{\widehat{D}}_n'(z).
    \end{align}
    To find $A$, we use the following normalisation condition:
    \begin{equation*}
        \mathcal{D}_n'(z=0) = 1 = A \, \mathcal{\widehat{D}}_n'(0),
    \end{equation*}
    which implies
    \begin{equation}
        \label{eq:A}
         A = \frac{1}{\mathcal{\widehat{D}}_n'(0)},
    \end{equation}
    so that we can finally construct the collections $\{\mathcal{D}_n\}$ and $\{\mathcal{D}'_n\}$ using eq.~\eqref{eq:collection_D} and \eqref{eq:collection_Dprime} without assuming any $H_0$ and while neutralizing any bias from the choice of $M_\mathrm{B}$

\section{DES~Y5 results}

\begin{figure}
        \centering
        \includegraphics[scale=1]{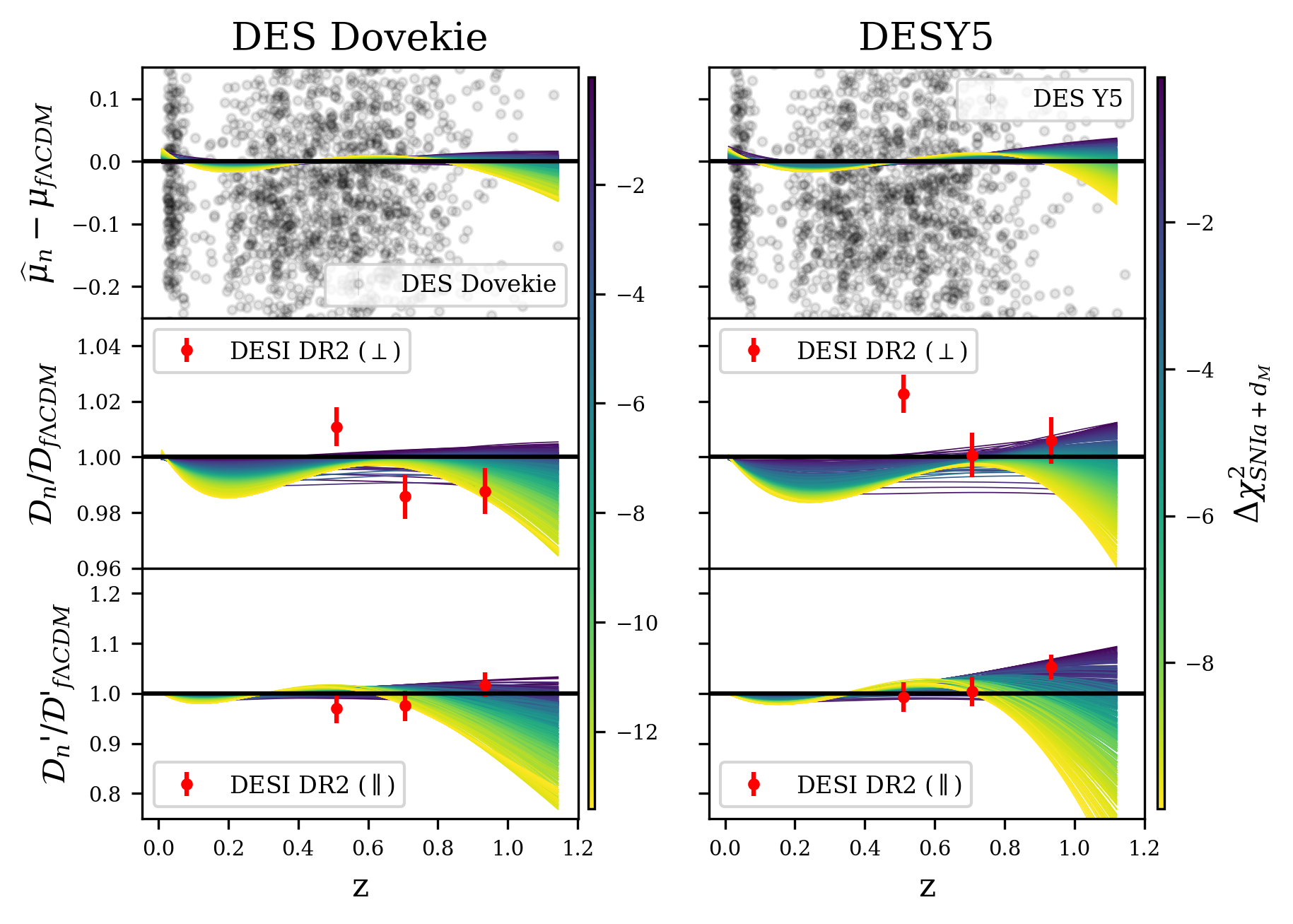}
        \caption{Comparison of the smoothing results between DES~Dovekie and DES~Y5}
        \label{fig:annex_smoothing}
    \end{figure}

\begin{figure}
        \centering
        \includegraphics[scale=1]{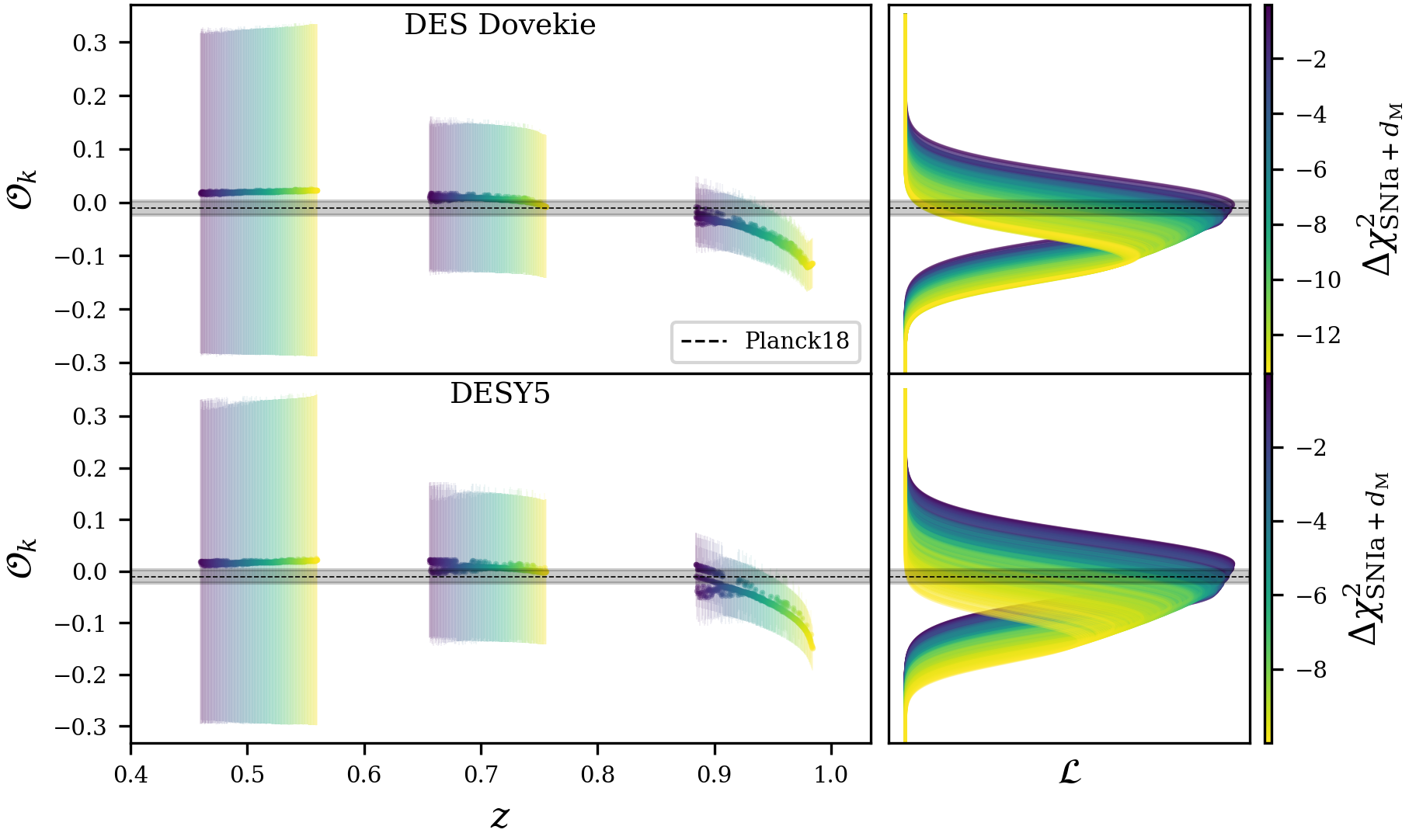}
        \caption{$\Ok$ diagnostic comparison between DES~Dovekie and DES~Y5, with selection criterion $\Delta \chi^2_{\mathrm{SNIa} + d_\mathrm{M}}<0$}
        \label{fig:annex_ok_diag}
    \end{figure}

    We compare the updated DES~Dovekie SNIa data with their predecessor DES~Y5. Figure \ref{fig:annex_smoothing} shows the smoothing results for both datasets. The selection criterion is $\Delta \chi^2_{\mathrm{SNIa} + d_\mathrm{M}}<0$. Although the differences are minimal, it seems that the DES~Dovekie reconstructions deviate slightly less from the best-fitting $\Lambda$CDM model than the DES~Y5 reconstructions. The associated $\Ok$ diagnostic is displayed on Figure \ref{fig:annex_ok_diag}, and the results for all selection criterion are recapitulated in Table \ref{tab:desy5_comparison}. We can see that the median curvature of DES~Y5 $\Omk[0]^\mathrm{med} = -0.116^{+0.131}_{-0.018} \pm 0.043$ is more negative than the median curvature of DES~Dovekie. Additionally, the maximum deviation from flatness among the reconstructed $\Ok$ diagnostic of DES~Y5 is -3.28$\sigma$, versus {-2.51$\sigma$} for DES~Dovekie, indicating that the updated data have reinforced the agreement with flatness. This is also visible on Figure \ref{fig:deviation}.
    
    \begin{table}
        \centering
        \begin{tabular}{c c c c}
        \toprule
        $\Delta\chi^2$ criterion 
        & Data combination 
        & ${\Omega}_{k,0}^\text{med}$ 
        & $\frac{c}{H_0 r_\mathrm{d}}$\\
        \midrule
        
        \multirow{2}{*}{$\Delta\chi^2_{\mathrm{SNIa}}<0$}
        & DES~Y5 \& DESI~DR2 
        & $-0.116^{+0.131}_{-0.018} \pm 0.043$ & $30.716^{+0.006}_{-0.422} \pm 0.161$\\
    
        & DES~Dov. \& DESI~DR2 
        & $-0.102^{+0.100}_{-0.005} \pm 0.043$ & $30.254^{+0.014}_{-0.422} \pm 0.159$\\
        \midrule
        
        \multirow{2}{*}{$\Delta\chi^2_{\mathrm{SNIa}+\dcom}<0$}
        & DES~Y5 \& DESI~DR2 
        & $-0.116^{+0.131}_{-0.018} \pm 0.043$ & $30.716^{+0.141}_{-0.409} \pm 0.161$\\
    
        & DES~Dov. \& DESI~DR2 
        & $-0.102^{+0.099}_{-0.005} \pm 0.043$ & $30.254^{+0.014}_{-0.380} \pm 0.159$\\
        
    
        \bottomrule
        \end{tabular}
        
        \caption{Comparison between the updated DES~Dovekie results and the original DES~Y5 data. The selection  criterion $\Delta \chi^2_{\mathrm{SNIa} + d_\mathrm{M}}<0$
        }
        \label{tab:desy5_comparison}
    \end{table}

\section{Smoothing scale}
\label{sec:smoothing_scale}

\begin{figure}
        \centering
        \includegraphics[scale=1.2]{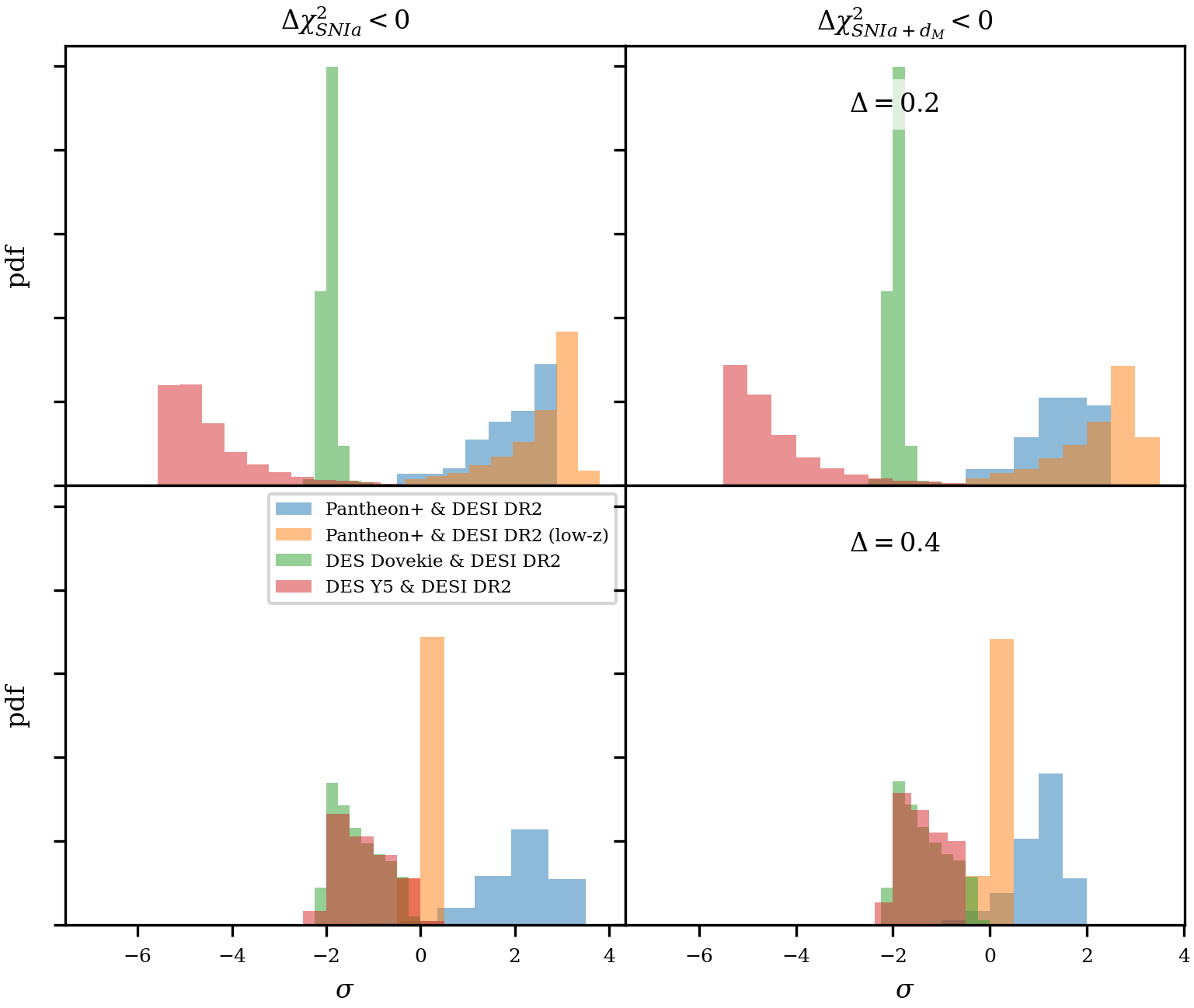}
        \caption{Histogram of the deviations from flatness for $\Delta = 0.2$ (top panel) and $\Delta=0.4$ (bottom panel).}
        \label{fig:hist_app}
    \end{figure}

Table \ref{tab:smoothing_scale} recapitulates the results of the $\Ok$ diagnostic after re-running the full pipeline with smoothing scale $\Delta = 0.2$ and $\Delta = 0.4$. The corresponding deviation histograms are in the displayed in the top and bottom rows of Fig.  \ref{fig:hist_app} respectively. 

With $\Delta = 0.2$, the smoothing is more sensitive to the noise and variations in the data. The peaks of the histograms in the top panel of Fig. \ref{fig:hist_app} are displaced toward stronger deviation compared to Fig. \ref{fig:deviation}, showing that the preference for negative $\Omk[0]$ in DES-Dovekie and DES~Y5, and positive $\Omk[0]$ in Pantheon+ have been reinforced by the smaller smoothing scale. In this case, the reconstructions from Pantheon+ \& DESI~DR2 (low-z) deviates more significantly from flatness than Pantheon+ \& DESI~DR2, which is not the case for higher values of $\Delta$. The Pantheon+ data become significantly sparser around $z\sim1$, and the reduction of the value of the smoothing scale generate stronger deviations toward positive values in the $\Ok$ diagnostic reconstructions at  $z=1.13$,  the maximum redshift of Pantheon+ \& DES~DR2 (low-z), and drive the peak of the histogram (in orange) toward high, positive values. Meanwhile, the two higher redshift data points, present in the Pantheon+ \& DESI~DR2 data combination (in blue) deviate into negative value, explaining the final weighted average to be closer to $\Ok=0$ in this case. 

With $\Delta=0.4$, the reconstructions are smoother and SNIa that strongly deviate from the rest of the sample have less impact on the results. As a results,  we can see in the bottom panel of Fig. \ref{fig:hist_app} that all reconstructions consistent with FLRW are also consistent with flatness within $3\sigma$ in absolute value. The previous preferences for different values of $\Omk[0]$ remain, but the higher smoothing scale attenuates the disagreement between dataset and increases the consistency with flatness.

\begin{table}
    \centering
    \begin{tabular}{c c c c}
    \toprule
     $\Delta \chi^2$ criterion & Data combination & $\Delta = 0.2$ & $\Delta = 0.4$ \\ 
    \midrule
    \multirow{3}{*}{$\Delta \chi^2_{\mathrm{SNIa}}<0$} & P+ \& DESI~DR2 & $0.083^{+0.039}_{-0.096} \pm 0.039$ & $0.089^{+0.042}_{-0.132} \pm 0.039$ \\ 
                          & P+ \& DESI~DR2 (low-z) & $0.209^{+0.065}_{-0.233} \pm 0.074$ & $0.011^{+0.016}_{-0.038} \pm 0.054$ \\ 
                          & DES~Dov. \& DESI~DR2 & $-0.089^{+0.085}_{-0.014} \pm 0.045$ & $-0.066^{+0.066}_{-0.027} \pm 0.047$ \\ \midrule
    \multirow{3}{*}{$\Delta \chi^2_{\mathrm{SNIa} + \dcom}<0$} & P+ \& DESI~DR2 & $0.059^{+0.043}_{-0.071} \pm 0.038$ & $0.041^{+0.035}_{-0.080} \pm 0.037$ \\ 
                          & P+ \& DESI~DR2 (low-z) & $0.179^{+0.075}_{-0.210} \pm 0.072$ & $0.011^{+0.016}_{-0.051} \pm 0.054$ \\ 
                           & DES~Dov. \& DESI~DR2 & $-0.089^{+0.082}_{-0.014} \pm 0.045$ & $-0.066^{+0.063}_{-0.027} \pm 0.047$ \\ 
    \bottomrule
    \end{tabular}

    \caption{Results of the $\Ok$ diagnostic for different values of the smoothing scale $\Delta$}
    \label{tab:smoothing_scale}
\end{table}

\acknowledgments

This work was partially supported by the ``PHC STAR'' program (project number: 50123WB, RS-2023-00259422), funded by the French Ministry for Europe and Foreign Affairs, the French Ministry for Higher Education and Research, and the National Research Foundation of Korea.
C.~M. acknowledges the support of the Center for the Gravitational Waves Universe (2021M3F7A1082056) and the National Research Foundation of Korea (NRF-2023R1A2C1006984).
B.~L. also acknowledges the support of the National Research Foundation of Korea (NRF-2022R1F1A1076338 \& NRF-RS-2024-00334550)
and the support of the Korea Institute for Advanced Study (KIAS) grant funded by the government of Korea.

\bibliographystyle{JHEP}
\bibliography{biblio.bib}

\end{document}